\documentclass{article}
\usepackage{amssymb}
\usepackage{amsfonts}
\usepackage{amsmath}

\input{tcilatex}

\begin{document}

\title{Elastic Plate Deformation with Transverse Variation of Microrotation }
\author{Lev Steinberg \\
Department of Mathematical Sciences\\
University of Puerto Rico\\
Mayaguez, PR 00681-9018, USA}
\maketitle

\begin{abstract}
The purpose of this paper is to present a new mathematical model for the
deformation of thin Cosserat elastic plates. Our approach, \ which is based
on a generalization of the classical Reissner plate theory, takes into
account the transverse variation of microrotation of the plates. The model
assumes polynomial approximations\ over the plate thickness of asymmetric
stress, couple stress, displacement, and microrotation, which are consistent
with the elastic equilibrium, boundary conditions and the constitutive
relationships. Based on the generalized Hellinger-Prange -Reissner
variational principle and strain-displacement relation we obtain the
complete theory of Cosserat plate. We also proved the solution uniqueness
for the plate boundary value problem. \ 

\smallskip


\end{abstract}

\section{Introduction}

The well known classical bending theory of elastic plates\ \cite{Love}, \cite%
{naghdi}, \cite{timoshenko}, was first presented by Kirchhoff in his thesis
(1850) and is described by a bi-harmonic differential equation \cite{Donell},%
\cite{timoshenko}. The usual assumption of this theory is that the normal to
the middle plane remains normal during deformation. Thus the theory neglects
transverse shear strain effects. A system of equations, which takes into
account the transverse shear deformation, has been developed by E. Reissner
(1945) \cite{Reissner}, \cite{Reissner 1}.

One of the advantages of Reissner's model is that it is able to determine
the reactions along the edges of a simply supported rectangular plate, where
classical theory leads to a concentrated reaction at the corners of the
plate. The Reissner theory has been applied to thin walled structures with
moderate thickness. The study of the relationships between these two models
has proved that the solution of the clamped Reissner plate approaches the
solution of the Kirchhoff plate as the thickness approaches zero \cite{Bathe}
and that the maximum bending can reach up to 20\% for moderate plate
thickness \cite{Donell}. The numerical calculations of bending behavior of
the plate of moderate thickness, \cite{Rossle} show high level agreement
between 3D \ and Reissner models. More remarks on the history of the
modeling of classic linear elastic plates can be found in \cite{Love}, \cite%
{Rossle}, \cite{Reissner 2}.

In order to describe deformation of elastic plates with microstructure that
possess grains, particles, fibers, and cellular structures \cite{Neff2}, 
\cite{Forest}. A. C. Eringen (1967) was the first to propose a theory of
plates in the framework of Cosserat (micropolar) Elasticity \cite{Eringen}.
His theory is based on a direct technique of integration of the Cosserat
Elasticity. The Eringen plate theory does not consider a transverse
variation of the microrotation over the thickness, which might be necessary
for rather thick plates under vertical load and pure twisting momentum. In
order to develop a theory of plates, which can be used for thin wall
structures with moderate thickness, we propose to use the classic Reissner
plate theory as a foundation for the modeling of Cosserat elastic plates.
Our approach, in addition to the traditional model, takes into account the
second order approximation of couple stresses and the variation of three
components of microrotation in the thickness direction.

\section{Micropolar (Cosserat) Linear Elasticity}

\subsection{Fundamental Equations}

Before proceeding some notation convention should be explained. We use the
usual summation conventions and all expressions that contain Latin letters
as subindices are understood to take values in the set $\{1,2,3\}$. When
Greek letters appear as subindices then it will be assumed that they can
take the values $1$ or $2$.

The Cosserat elasticity equilibrium equations without body forces represent
the balance of linear and angular momentums of micropolar elasticity and
have the following form \cite{Eringen}: 
\begin{eqnarray}
\mathbf{div\sigma }\mathbf{=0} &&,  \label{equilibrium_equations 0} \\
\mathbf{\varepsilon \cdot \sigma }\text{ +}\mathbf{div\mu } &\mathbf{=}&%
\mathbf{0},  \label{equilibrium_equations}
\end{eqnarray}%
{where} the quantity $\mathbf{\sigma =}\left\{ \sigma _{ji}\right\} $ is the
stress tensor, $\mathbf{\mu =}\left\{ \mu _{ji}\right\} $ the couple stress
tensor, $\mathbf{\varepsilon }=\left\{ \varepsilon _{ijk}\right\} $ is the
Levi--Civita tensor, where $\varepsilon _{ijk}$ equals 1 or -1 according as $%
(i,j,k)$ is an even or odd permutation of 1,2,3 and zero otherwise, and $%
\mathbf{\varepsilon \cdot \sigma =}\left\{ \mathbf{\varepsilon }_{ijk}\sigma
_{jk}\right\} .$

The constitutive equations can be written in the form \cite{Nowaki}:

\begin{eqnarray}
\mathbf{\sigma } &=&(\mu +\mu _{c})\mathbf{\gamma }+(\mu -\mu _{c})\mathbf{%
\gamma }^{T}+\lambda \mathbf{(tr\gamma )1},  \label{Hooke's_law 1} \\
\mathbf{\mu } &=&(\gamma +\epsilon )\mathbf{\chi }+(\gamma -\epsilon )%
\mathbf{\chi }^{T}+\beta \mathbf{(tr\chi )1},  \label{Hooke's_law 1A}
\end{eqnarray}%
{and the strain-displacement and torsion-rotation relations}

\begin{equation}
\text{ }\mathbf{\gamma =}\left( \mathbf{\nabla u}\right) ^{T}\mathbf{%
+\varepsilon \cdot \varphi }\text{ and }\mathbf{\chi }=\nabla \mathbf{%
\varphi },  \label{kinematic formulas}
\end{equation}%
where quantities $\mathbf{\gamma }$ and $\mathbf{\chi }$ $,$ are the
micropolar strain and torsion tensors,\textbf{\ }$\mathbf{u}$ and \textbf{\ }%
$\mathbf{\varphi }$ \ the displacement and rotation vectors respectively, $%
\mathbf{1}$ the identity tensor, and $\mathbf{\mu },\lambda $ are the
symmetric, $\mu _{c},\beta \mathbf{,}\gamma \mathbf{,}\epsilon $ the
asymmetric Cosserat elasticity constants.

In the reversible form:%
\begin{eqnarray}
\mathbf{\gamma } &\mathbf{=}&(\mu ^{\prime }+\mu _{c}^{\prime })\mathbf{%
\sigma }+(\mu ^{\prime }-\mu _{c}^{\prime })\mathbf{\sigma }^{T}+\lambda
^{\prime }\mathbf{(tr\sigma )1},  \label{Hooke's_law 2} \\
\mathbf{\chi } &\mathbf{=}&(\gamma ^{\prime }+\epsilon ^{\prime })\mathbf{%
\mu }+(\gamma ^{\prime }-\epsilon ^{\prime })\mathbf{\mu }^{T}+\beta \mathbf{%
(tr\mu )1}.  \label{Hooke's_law 2A}
\end{eqnarray}%
where{\ }$\mu ^{\prime }=\frac{1}{4\mu }$, $\mu _{c}^{\prime }=\frac{1}{4\mu
_{c}}$, $\gamma ^{\prime }=\frac{1}{4\gamma }$, $\epsilon ^{\prime }=\frac{1%
}{4\epsilon }$, $\lambda ^{\prime }=\frac{-\lambda }{6\mu (\lambda +\frac{%
2\mu }{3})}$ and $\beta ^{\prime }=\frac{-\beta }{6\mu (\beta +\frac{2\gamma 
}{3})}$.

We consider a Cosserat elastic body $B_{0}.$ In this case the equilibrium
equations (\ref{equilibrium_equations 0}) - (\ref{equilibrium_equations})
with constitutive formulas (\ref{Hooke's_law 1}) - (\ref{Hooke's_law 1A})
and kinematics formulas (\ref{kinematic formulas}) should be accompanied by
the following mixed boundary conditions%
\begin{eqnarray}
\mathbf{u} &\mathbf{=}&\mathbf{u}_{o}\mathbf{,}\text{ }\mathbf{\varphi
=\varphi }_{o}\text{ on }\mathcal{G}_{1}=\partial B_{0}\backslash \partial
B_{\sigma },  \label{Bound conditions} \\
\mathbf{\sigma }_{\mathbf{n}} &=&\mathbf{\sigma \cdot n}=\mathbf{\sigma }%
_{o},\text{ }\mathbf{\mu }_{\mathbf{n}}=\mathbf{\mu }\cdot \mathbf{n=\mu }%
_{o}\text{ on }\mathcal{G}_{2}=\partial B_{\sigma },
\label{Bound conditions a}
\end{eqnarray}%
where $\mathbf{u}_{o}$\textbf{, }$\mathbf{\varphi }_{o}$\textbf{\ }are
prescribed on $\mathcal{G}_{1}$, $\sigma _{o}$\textbf{\ }and \ $\mathbf{\mu }
$ \ on $\mathcal{G}_{2},$ and $\mathbf{n}$\textbf{\ }denotes the outward
unit normal vector to $\partial B_{0}.$

\subsection{Cosserat Elastic Energy}

The strain stored energy $U_{C}$ of the body $B_{0}$ is defined by the
integral \cite{Nowaki}:

\begin{equation}
{\large U}_{C}=\int_{B_{0}}\emph{W}\left\{ \mathbf{\gamma ,\chi }\right\} dv,
\label{free_energy_strain}
\end{equation}%
where 
\begin{eqnarray}
\emph{W}\left\{ \mathbf{\gamma ,\chi }\right\} &=&\frac{\mu +\mu _{c}}{2}%
\gamma _{ij}\gamma _{ij}+\frac{\mu -\mu _{c}}{2}\gamma _{ij}\gamma _{ji}+%
\frac{\lambda }{2}\gamma _{kk}\gamma _{nn}  \label{free_energy_strain A} \\
&&+\frac{\gamma +\epsilon }{2}\chi _{ij}\chi _{ij}+\frac{\gamma -\epsilon }{2%
}\chi _{ij}\chi _{ji}+\frac{\beta }{2}\chi _{kk}\chi _{nn},  \notag
\end{eqnarray}%
then the constitutive relations (\ref{Hooke's_law 1}) - (\ref{Hooke's_law 1A}%
) can be written in the form:

\begin{equation}
\mathbf{\sigma }=\mathbf{C}_{\sigma }\left[ \emph{W}\right] =\nabla _{%
\mathbf{\gamma }}\emph{W}\text{ and }\mathbf{\mu }=\mathbf{C}_{\mu }\left[ 
\emph{W}\right] =\nabla _{\mathbf{\chi }}\emph{W}.  \label{Const-d 2}
\end{equation}

The function $\emph{W}$ is positive if and only if \cite{Nowaki} 
\begin{eqnarray}
\mu &>&0,\ \ 3\lambda +2\mu >0,  \notag \\
\gamma &>&0,\ \ 3\beta +2\gamma >0,  \label{parameter conditions} \\
\mu _{c} &>&0,\ \ \mu +\mu _{c}>0,\text{ }\epsilon >0.  \notag
\end{eqnarray}%
The following conditions \cite{Neff 1} for the Cosserat elastic energy%
\begin{eqnarray}
\mu &>&0,\ \ 3\lambda +2\mu >0,  \notag \\
\gamma &>&0,\ \ 3\beta +2\gamma >0, \\
\mu _{c} &\geq &0,\ \ \epsilon \geq 0.  \notag
\end{eqnarray}%
are enough to provide the uniqueness of static problems.

For future convenience, we present the stress energy%
\begin{equation*}
{\large U}_{K}=\int_{B_{0}}\Phi \left\{ \sigma \mathbf{,\mu }\right\} dv,
\end{equation*}%
where%
\begin{eqnarray}
\Phi \left\{ \sigma \mathbf{,\mu }\right\} &=&\frac{\mu ^{\prime }+\mu
_{c}^{\prime }}{2}\sigma _{ij}\sigma _{ij}+\frac{\mu ^{\prime }-\mu
_{c}^{\prime }}{2}\sigma _{ij}\sigma _{ji}+\frac{\lambda ^{\prime }}{2}%
\sigma _{kk}\sigma _{nn}  \notag \\
&&+\frac{\gamma ^{\prime }+\epsilon ^{\prime }}{2}\mu _{ij}\mu _{ij}+\frac{%
\gamma ^{\prime }-\epsilon ^{\prime }}{2}\mu _{ij}\mu _{ji}+\frac{\beta
^{\prime }}{2}\mu _{kk}\mu _{nn}.  \label{free_energy_stress}
\end{eqnarray}%
\ The reversible constitutive relation (\ref{Hooke's_law 2}) - (\ref%
{Hooke's_law 2A}) can be also written in form:%
\begin{equation}
\mathbf{\gamma =K}_{\gamma }\left[ \mathbf{\sigma }\right] =\frac{\partial
\Phi }{\partial \mathbf{\sigma }},\text{ }\mathbf{\chi =K}_{\chi }\left[ 
\mathbf{\mu }\right] =\frac{\partial \Phi }{\partial \mathbf{\mu }}.
\label{Const1-1B}
\end{equation}

The total internal work done by the stresses $\mathbf{\sigma }$ and $\mathbf{%
\mu }$ over the strains $\mathbf{\ \gamma }$ and $\mathbf{\chi }$ $\ $for
the body $B_{0}$ \cite{Nowaki} is 
\begin{equation}
{\large U}=\int_{B_{0}}\left[ \mathbf{\sigma \cdot \gamma +\mu \cdot \chi }%
\right] dv  \label{free energy}
\end{equation}%
and 
\begin{equation*}
{\large U=U}_{K}{\large =U}_{C}
\end{equation*}%
provided the constitutive relations (\ref{Hooke's_law 1}) - (\ref%
{Hooke's_law 1A}) hold.

\subsection{The Generalized Hellinger-Prange -Reissner (HPR)\ Principle}

The HPR principle \cite{Gurtin} in the case of Cosserat elasticity\ states,
that for any set $\mathcal{A}$ of all admissible states $\mathfrak{s=}\left[ 
\mathbf{u,\varphi ,\gamma ,\chi ,\sigma ,\mu }\right] $ that satisfy the {%
strain-displacement and torsion-rotation relations (\ref{kinematic formulas}%
),} the zero variation 
\begin{equation*}
\delta \Theta (\mathfrak{s})=0
\end{equation*}

of the functional 
\begin{eqnarray}
\Theta (\mathfrak{s}) &=&U_{K}-\int_{B_{0}}\left[ \mathbf{\sigma \cdot
\gamma +\mu \cdot \chi }\right] dv  \label{Var 2} \\
&&+\int_{\mathcal{G}_{1}}\left[ \mathbf{\sigma }_{\mathbf{n}}\cdot (\mathbf{%
u-u}_{o})+\mathbf{\mu }_{\mathbf{n}}\left( \mathbf{\varphi -\varphi }%
_{o}\right) \right] da+\int_{\mathcal{G}_{2}}\left[ \mathbf{\sigma }%
_{o}\cdot \mathbf{u+m}_{o}\cdot \mathbf{\varphi }\right] da  \notag
\end{eqnarray}%
\bigskip at $\mathfrak{s\in }\mathcal{A}$ is equivalent of $\mathfrak{s}$ to
be a solution of the system of equilibrium equations (\ref%
{equilibrium_equations 0}) - (\ref{equilibrium_equations}), constitutive
relations (\ref{Hooke's_law 2}) - (\ref{Hooke's_law 2A}), which satisfies
the mixed boundary conditions (\ref{Bound conditions}) - (\ref{Bound
conditions a}). The proof is similar to the proof for HPR\ principle for
classic linear elasticity \cite{Gurtin}.

\section{The Cosserat Plate Assumptions}

In this section we formulate our stress, couple stress and kinematic
assumptions of the Cosserat plate. The set of points $P=\left\{ \Gamma
\times \left[ -h/2,h/2\right] \right\} \cup T\cup B$ forms the entire
surface of the plate and $\left\{ \Gamma _{u}\times \left[ -h/2,h/2\right]
\right\} $ is the lateral part of the boundary where displacements and
microrotations are prescribed. The notation $\Gamma _{\sigma }=\Gamma
\backslash \Gamma _{u}$ of the remainder we use to describe the lateral part
of the boundary edge $\left\{ \Gamma _{\sigma }\times \left[ -h/2,h/2\right]
\right\} $ where stress and couple stress are prescribed. We also use
notation $P_{0}$ \ for the middle plane internal domain of the plate.

In our case we consider the vertical load and pure twisting momentum
boundary conditions at the top and bottom of the plate, which can be written
in the form:

\begin{eqnarray}
\sigma _{33}(x_{1},x_{2},h/2) &=&\sigma ^{t}(x_{1},x_{2}),\text{ }\sigma
_{33}(x_{1},x_{2},-h/2)=\sigma ^{b}(x_{1},x_{2}),  \label{Bound conditions 0}
\\
\text{ \ \ \ \ \ }\sigma _{3\beta }(x_{1},x_{2},\pm h/2) &=&0,
\label{Bound conditions 1} \\
\text{ \ \ \ }\mu _{33}(x_{1},x_{2},h/2) &=&\mu ^{t}(x_{1},x_{2}),\text{ }%
\mu _{33}(x_{1},x_{2},-h/2)=\mu ^{b}(x_{1},x_{2}),
\label{Bound conditions 1a} \\
\text{\ }\mu _{3\beta }(x_{1},x_{2},\pm h/2) &=&0,
\label{Bound conditions 2}
\end{eqnarray}%
where $(x_{1},x_{2})\in P_{0}.$

\subsection{Stress and Couple Stress Assumptions}

Our approach, which is in the spirit of the Reissner's theory of plates \cite%
{Reissner}, assumes that the variation of stress $\sigma _{kl\text{ }}$and
couple stress $\mu _{kl}$ components across the thickness can be represented
by means of polynomials of $x_{3}$ in such a way that it will be consistent
with the equilibrium equations (\ref{equilibrium_equations 0}) and (\ref%
{equilibrium_equations}). First, as it is assumed in the standard theory of
plates, we use expressions for the stress components in the following form 
\cite{Wan}:

\begin{equation}
\sigma _{\alpha \beta }=n_{\alpha \beta }(x_{1},x_{2})+\frac{h}{2}\zeta
_{3}m_{\alpha \beta }(x_{1},x_{2}),  \label{stress assumption 1}
\end{equation}%
where $\zeta _{3}=\frac{2}{h}x_{3},\ $\ and $\alpha ,\beta \in \{1,2\}$. The
only difference between our assumptions and those of Reissner' \cite%
{Reissner} is that the functions $n_{\alpha \beta }$ and $m_{\alpha \beta }$
are not\ symmetric. Based on \ (\ref{stress assumption 1}) and by means of
the first two equations of written in the component form stress equilibrium (%
\ref{equilibrium_equations 0}) 
\begin{equation*}
\sigma _{j\beta ,j}=0
\end{equation*}%
we obtain for the shear stress components

\begin{equation}
\sigma _{3\beta }=q_{\beta }(x_{1},x_{2})\left( 1-\zeta _{3}^{2}\right) ,
\label{stress assumption 1a}
\end{equation}

It is natural to assume that the expressions for the remaining shear stress
component are in the form similar to (\ref{stress assumption 1a}), i.e.

\begin{equation}
\sigma _{\beta 3}=q_{\beta }^{\ast }(x_{1},x_{2})\left( 1-\zeta
_{3}^{2}\right) .  \label{stress assumption 1b}
\end{equation}%
Here, as is usual for the asymmetric elasticity, the functions $q_{\beta
},q_{\beta }^{\ast }.$ can be different.

Substituting equations (\ref{stress assumption 1b}) in the remaining
equilibrium differential equation for stress%
\begin{equation*}
\sigma _{j3,j}=0\ 
\end{equation*}%
we obtain the expression for the transverse normal stress

\begin{equation}
\sigma _{33}=\zeta _{3}\left( \frac{1}{3}\zeta _{3}^{2}-1\right) k^{\ast
}(x_{1},x_{2})+m^{\ast }(x_{1},x_{2}).  \label{stress assumption 1c}
\end{equation}

The next step is to accommodate approximations (\ref{stress assumption 1c})
to the boundary conditions (\ref{Bound conditions 0}). By direct
substitution to (\ref{Bound conditions 0}) it easy to obtain that 
\begin{equation}
\sigma _{33}=-\frac{3}{4}\left( \frac{1}{3}\zeta _{3}^{3}-\zeta _{3}\right)
p+\sigma _{0},  \label{stress assumption 1ca}
\end{equation}%
where $p=\sigma ^{t}(x_{1},x_{2})-\sigma ^{b}(x_{1},x_{2})$ and $\sigma _{0}=%
\frac{1}{2}\left( \sigma ^{t}(x_{1},x_{2})+\sigma ^{b}(x_{1},x_{2})\right) $
satisfy the boundary condition requirements. We note that expression (\ref%
{stress assumption 1ca}) is identical to the expression of $\sigma _{33}$
given in \cite{Reissner} in the case of $\sigma ^{b}=0$.

It is also assumed that the couple stress $\mu _{\alpha \beta }$ should have
expression similar to the shear stress $\sigma _{3\beta }$\ expressions (\ref%
{stress assumption 1a}) - (\ref{stress assumption 1b}):%
\begin{equation}
\mu _{\alpha \beta }=\left( 1-\zeta _{3}^{2}\right) r_{\alpha \beta
}(x_{1},x_{2}).  \label{stress assumption 1d}
\end{equation}%
Finally we assume that couple stress $\mu _{\beta 3}$ expression is similar
to $\sigma _{\alpha \beta }$ (\ref{stress assumption 1}):

\begin{equation}
\mu _{\beta 3}=\zeta _{3}s_{\beta }^{\ast }(x_{1},x_{2})+m_{\beta }^{\ast
}(x_{1},x_{2}).  \label{stress assumption 1e}
\end{equation}%
Note that the first two equations of (\ref{equilibrium_equations})\ can be
written in the form 
\begin{equation}
\epsilon _{\beta jk}\sigma _{jk}+\mu _{j\beta ,j}=0,  \label{eq22}
\end{equation}%
and substituting the couple stress (\ref{stress assumption 1d}) in (\ref%
{eq22}) and taking into account (\ref{stress assumption 1a}) and (\ref%
{stress assumption 1b}) we obtain the expression for the transverse shear
couple stress: 
\begin{equation}
\mu _{3\beta }=\left( \frac{1}{3}\zeta _{3}^{3}-\zeta _{3}\right) s_{\beta
}(x_{1},x_{2})+m_{\beta }(x_{1},x_{2}).  \label{couple stress 1}
\end{equation}%
Substituting (\ref{couple stress 1}) to boundary conditions (\ref{Bound
conditions 1a}) we obtain that

\begin{equation*}
s_{\beta }(x_{1},x_{2})=0\text{ and }m_{\beta }(x_{1},x_{2})=0,
\end{equation*}%
i.e. the transverse shear couple stress 
\begin{equation}
\mu _{3\beta }=0.  \label{couple stress 1a}
\end{equation}%
Now, substituting the couple stress (\ref{stress assumption 1e})\ and stress
(\ref{stress assumption 1}) in the remaining differential equation of the
equilibrium of angular momentum (\ref{equilibrium_equations}) 
\begin{equation}
\epsilon _{3jk}\sigma _{jk}+\mu _{j3,j}=0,  \label{eq33}
\end{equation}%
we obtain the transverse normal couple stress to be in the form: 
\begin{equation}
\mu _{33}=\frac{1}{2}\zeta _{3}^{2}a^{\ast }(x_{1},x_{2})+\zeta _{3}b^{\ast
}(x_{1},x_{2})+c^{\ast }(x_{1},x_{2}).  \label{couple stress 2}
\end{equation}%
The next step is to accommodate boundary conditions (\ref{Bound conditions
1a}) to (\ref{couple stress 2}). At this stage we restrict \ the form of (%
\ref{couple stress 2}), which could allow us to determine couple stress $\mu
_{33}$ \ directly from boundary conditions (\ref{Bound conditions 1a}). To
this end we make an additional assumption that $\mu _{33}$ must be a first
order polynomial%
\begin{equation}
\mu _{33}=\zeta _{3}b^{\ast }(x_{1},x_{2})+c^{\ast }(x_{1},x_{2}).
\label{couple stress 2a}
\end{equation}%
This assumption is also consistent with the equilibrium equation (\ref{eq33}%
) and allows us to proceed as we did for the determination of transverse
loading stress (\ref{stress assumption 1ca}) from the stress boundary
conditions. Now boundary conditions\ (\ref{Bound conditions 1a}) are
sufficient to determine $\mu _{33}$, which must be of the form%
\begin{equation}
\mu _{33}=\zeta _{3}v+t,\   \label{stress assump 3(a)}
\end{equation}%
where $v(x_{1},x_{2})=\frac{1}{2}\left( \mu ^{t}(x_{1},x_{2})-\mu
^{b}(x_{1},x_{2})\right) $ and $t(x_{1},x_{2})=\frac{1}{2}\left( \mu
^{t}(x_{1},x_{2})+\mu ^{b}(x_{1},x_{2})\right) $.

\subsection{Kinematic Assumptions}

The choice of kinematic assumptions\ is based on simplicity and their
compatibility with the constitutive relationships of stress and couple
stress assumptions (\ref{Hooke's_law 1}). As in the standard theory of thin
plates, it is assumed that \ $u_{a}$ displacements are distributed linearly
over the thickness of the plate \cite{Eringen}\ and \ that $u_{3}$ \ does
not vary over the thickness of the plate, i.e.

\begin{eqnarray}
u_{\alpha } &=&U_{\alpha }(x_{1},x_{2})-\frac{h}{2}\zeta _{3}V_{\alpha
}(x_{1},x_{2}),  \label{kin 1} \\
u_{3} &=&w(x_{1},x_{2}),  \notag
\end{eqnarray}%
The terms $V_{\alpha }(x_{1},x_{2})$ in (\ref{kin 1}) represent the
rotations in middle plane.

In order to accommodate the transverse microrotations to the constitutive
relations\ (\ref{Hooke's_law 1}) we propose the variation of microrotation
with respect to $x_{3}$ by means of the second and third order polynomials$:$

\begin{eqnarray}
\varphi _{\alpha } &=&\Theta _{\alpha }^{0}(x_{1},x_{2})\left( 1-\zeta
_{3}^{2}\right) ,  \label{kin2a} \\
\varphi _{3} &=&\Theta _{3}^{0}(x_{1},x_{2})+\zeta _{3}\left( 1-\frac{1}{3}%
\zeta _{3}^{2}\right) \Theta _{3}(x_{1},x_{2}).  \label{kin 2}
\end{eqnarray}%
The constitutive formulas (\ref{Hooke's_law 1}) - (\ref{Hooke's_law 1A})
motivate us to chose the forms (\ref{kin2a}) and (\ref{kin 2}), which
produce expressions for $\varphi _{\alpha ,\beta }$ and $\varphi _{3,3}$
similar to what we have for couple stress approximations (\ref{stress
assumption 1d}).

{The functions \ } $\Theta _{i}^{0}$ \ in (\ref{kin2a}) and (\ref{kin 2})
describe microrotation components in the middle plane of the plate and $%
\Theta _{3}(x_{1},x_{2})$ the slope at the middle plane. Thus, in the
assumptions (\ref{kin2a}) and (\ref{kin 2}) the transverse variation effect
of microrotations is not neglected.

\section{Specification of HPR\ Variational Principle for the Cosserat Plate}

The HPR\ variational principle for a Cosserat plate is most appropriately
expressed in terms of corresponding integrands calculated across the whole
thickness. We also introduce the weighted characteristics of displacements,
microrotations, strains and stresses of the plate, which will be used to
produce the explicit forms of these integrands.

\subsection{The Cosserat plate stress energy density}

We define the plate stress energy density by the formula;

\begin{equation}
{\large \Phi (}\mathcal{S}{\large )}=\frac{h}{2}\int_{-1}^{1}\Phi \left\{
\sigma \mathbf{,\mu }\right\} d\zeta _{3}.  \label{C and Co expression}
\end{equation}

Taking into account the stress and couple stress assumptions (\ref{stress
assumption 1}) - (\ref{stress assump 3(a)}) and by the integrating $\Phi
\left\{ \sigma \mathbf{,\mu }\right\} $ with respect $\zeta _{3\text{ }}$in $%
[-1,1]$ \ we obtain the explicit plate stress energy density expression in
the form:

\begin{eqnarray}
{\large \Phi (}\mathcal{S}{\large )} &=&\frac{\lambda +\mu }{2h\mu (3\lambda
+2\mu )}\left[ N_{\alpha \alpha }^{2}+\frac{12}{h^{2}}M_{\alpha \alpha }^{2}%
\right]   \notag \\
&&-\frac{\lambda }{2h\mu (3\lambda +2\mu )}\left[ N_{11}N_{22}+\frac{12}{%
h^{2}}M_{11}M_{22}\right]   \notag \\
&&+\frac{\mu _{c}+\mu }{8h\mu _{c}\mu }\left[ (1-\delta _{\alpha \beta
})\left( N_{\alpha \beta }^{2}+\frac{12}{h^{2}}M_{\alpha \beta }^{2}\right) +%
\frac{6}{5}\left( Q_{\alpha }Q_{\alpha }+Q_{\beta }^{\ast }Q_{\beta }^{\ast
}\right) \right]   \notag \\
&&+\frac{3(\mu _{c}-\mu )}{10h\mu _{c}\mu }\left[ Q_{\alpha }Q_{\alpha
}^{\ast }+\frac{5}{6}N_{12}N_{21}+\frac{10}{h^{2}}M_{12}M_{21}\right] -\frac{%
3\lambda }{5h\mu (3\lambda +2\mu )}pM_{\beta \beta }  \notag \\
&&+\frac{3}{5h\gamma (3\beta +2\gamma )}\left[ (\beta +\gamma )R_{\alpha
\alpha }^{2}-\beta R_{11}R_{22}\right] +\frac{3}{10h}\left( \frac{1}{\gamma }%
-\frac{1}{\epsilon }\right) R_{12}R_{21}  \notag \\
&&+\frac{17h(\lambda +\mu )}{280\mu (3\lambda +2\mu )}p^{2}-\frac{\lambda }{%
2\mu (3\lambda +2\mu )}\left( N_{\alpha \alpha }\right) \sigma _{0}  \notag
\\
&&+\frac{h(\lambda +\mu )}{2\mu (3\lambda +2\mu )}\sigma _{0}^{2}-\frac{%
\gamma +\epsilon }{h\gamma \epsilon }\left[ \frac{1}{8}M_{\alpha }^{\ast
}M_{\alpha }^{\ast }+\frac{3}{2h^{2}}S_{\alpha }^{\ast }S_{\alpha }^{\ast }+%
\frac{3}{20}(1-\delta _{\beta \gamma })R_{\beta \gamma }^{2}\right]   \notag
\\
&&-\frac{\beta }{2\gamma (3\beta +2\gamma )}R_{\alpha \alpha }t+\frac{%
h(\beta +\gamma )}{2\gamma (3\beta +2\gamma )}t^{2}+\frac{h(\beta +\gamma )}{%
6\gamma (3\beta +2\gamma )}v^{2},  \label{Energy density}
\end{eqnarray}%
{where the Cosserat stress set }%
\begin{equation}
\mathcal{S}=\left[ M_{\alpha \beta },Q_{\alpha },Q_{3\alpha }^{\ast
},R_{\alpha \beta },S_{\beta }^{\ast },N_{\alpha \beta },M_{\alpha }^{\ast }%
\right] ,  \label{S1}
\end{equation}%
where

\begin{eqnarray}
M_{\alpha \beta } &=&\left( \frac{h}{2}\right) ^{2}\int_{-1}^{1}\zeta
_{3}\sigma _{\alpha \beta }d\zeta _{3}=\frac{h^{3}}{12}m_{\alpha \beta },
\label{Stress resultant 1} \\
Q_{\alpha } &=&\frac{h}{2}\int_{-1}^{1}\sigma _{3\alpha }d\zeta _{3}=\frac{2h%
}{3}q_{\alpha },\text{ }Q_{\alpha }^{\ast }=\frac{h}{2}\int_{-1}^{1}\sigma
_{\alpha 3}d\zeta _{3}=\frac{2h}{3}q_{\alpha }^{\ast }  \notag \\
R_{\alpha \beta } &=&\frac{h}{2}\int_{-1}^{1}\mu _{\alpha \beta }d\zeta _{3}=%
\frac{2h}{3}r_{\alpha \beta },  \notag \\
S_{\alpha }^{\ast } &=&\left( \frac{h}{2}\right) ^{2}\int_{-1}^{1}\zeta
_{3}\mu _{\alpha 3}d\zeta _{3}=\frac{h^{2}}{6}s_{\alpha }^{\ast },  \notag \\
N_{\alpha \beta } &=&\frac{h}{2}\int_{-1}^{1}\sigma _{\alpha \beta }d\zeta
_{3}=hn_{\alpha \beta },\text{ }M_{\alpha }^{\ast }=\frac{h}{2}%
\int_{-1}^{1}\mu _{\alpha 3}d\zeta _{3}=hm_{\alpha }^{\ast },  \notag
\end{eqnarray}

Here $M_{11}$ and $M_{22}$ are the bending moments, $M_{12}$ and $M_{21}$
the twisting moments, $Q_{\alpha }$ the shear forces, $Q_{\alpha }^{\ast }$
the transverse shear forces, $R_{11}$ and $R_{22}$ the micropolar bending
moments, $R_{12}$ and $R_{21}$ the micropolar twisting moments, $S_{\alpha
}^{\ast }$ the micropolar couple moments, all defined per unit length, $%
N_{11}$ and $N_{22}$ are the bending forces, $N_{12}$ and $N_{21}$ the
twisting forces, $M_{\alpha }^{\ast }$ the micropolar shear couple-stress
resultants.

Then the stress energy of the plate $P$%
\begin{equation}
U_{K}^{\mathcal{S}}=\int_{P_{0}}{\large \Phi (}\mathcal{S}{\large )}da,
\label{energy 1}
\end{equation}%
where $P_{0\text{ }}$ is the internal domain of the middle plane of the
plate $P.$

\subsection{The density of the work done over the Cosserat plate boundary}

In the following consideration we also assume that the proposed stress,
couple stress, and kinematic assumptions are valid for the lateral\ boundary
of the plate $P$ as well.

We evaluate the density of the work over the boundary $\Gamma _{u}\times %
\left[ -h/2,h/2\right] $

\begin{equation}
\mathcal{W}_{1}=\frac{h}{2}\int_{-1}^{1}\left[ \mathbf{\sigma }_{\mathbf{n}%
}\cdot \mathbf{u}+\mathbf{\mu }_{\mathbf{n}}\mathbf{\varphi }\right] d\zeta
_{3}.  \label{work 1}
\end{equation}%
Taking into account the stress and couple stress assumptions (\ref{stress
assumption 1}) - (\ref{stress assump 3(a)}) and kinematic assumptions (\ref%
{kin 1}) - (\ref{kin 2}) we are able to represent $\mathcal{W}_{1}$ by the
following expression: 
\begin{equation}
\mathcal{W}_{1}=\mathcal{S}_{n}\mathcal{\cdot U=}\check{M}_{\alpha }\Psi
_{\alpha }+\check{Q}^{\ast }W+\check{R}_{\alpha }\Omega _{\alpha }^{0}+%
\check{S}^{\ast }\Omega _{3}+\check{N}_{\alpha }U_{\alpha }+\check{M}^{\ast
}\Omega _{3}^{0},
\end{equation}%
where the sets $\mathcal{S}_{n}\mathcal{\ }$and $\mathcal{U}$ are defined as%
\begin{eqnarray*}
\mathcal{S}_{n} &=&\left[ \check{M}_{\alpha },\check{Q}^{\ast },\check{R}%
_{\alpha },\check{S}^{\ast },\check{N}_{\alpha },\check{M}^{\ast }\right] ,
\\
\mathcal{U} &=&\left[ \Psi _{\alpha },W,\Omega _{\alpha }^{0},\Omega
_{3},U_{\alpha },\Omega _{3}^{0}\right]
\end{eqnarray*}%
and

\begin{eqnarray*}
\check{M}_{\alpha } &=&M_{\alpha \beta }n_{\beta },\text{ }\check{Q}^{\ast
}=Q_{\beta }^{\ast }n_{\beta },\text{ }\check{R}_{\alpha }=R_{\alpha \beta
}n_{\beta }, \\
\check{S}^{\ast } &=&S_{\beta }^{\ast }n_{\beta },\text{ }\check{N}_{\alpha
}=N_{\alpha \beta }n_{\beta },\check{M}^{\ast }=M_{\beta }^{\ast }n_{\beta },
\end{eqnarray*}%
In the above $n_{\beta }$ is the outward unit normal vector to $\Gamma _{u},$
and

\begin{eqnarray}
\Psi _{\alpha } &=&\frac{3}{h}\int_{-1}^{1}\zeta _{3}u_{\alpha }d\zeta _{3},
\notag \\
W &=&\frac{3}{4}\int_{-1}^{1}\left( 1-\zeta ^{2}\right) u_{3}d\zeta _{3}, 
\notag \\
\Omega _{\alpha }^{0} &=&\frac{3}{4}\int_{-1}^{1}\left( 1-\zeta ^{2}\right)
\varphi _{\alpha }d\zeta _{3}  \label{lagrange multipliers} \\
\Omega _{3} &=&\frac{3}{h}\int_{-1}^{1}\zeta _{3}\varphi _{3}d\zeta _{3} 
\notag \\
U_{\alpha } &=&\frac{1}{2}\int_{-1}^{1}u_{\alpha }d\zeta _{3},  \notag \\
\Omega _{3}^{0} &=&\frac{1}{2}\int_{-1}^{1}\varphi _{3}d\zeta _{3},  \notag
\end{eqnarray}%
Here $\Psi _{\alpha }$ are the rotations of the middle plane around $%
x_{\alpha }$ axis, $W$ \ the vertical deflection of the middle plate, $%
\Omega _{k}^{0}$ the microrotations in the middle plate around $x_{k}$ axis$%
, $ $U_{\alpha }$ is the in-plane displacements of the middle plane along $%
x_{a}$ axis$,$ $\Omega _{3}$ the rate of change of the microrotation $%
\varphi _{3}$ along $x_{3}$.

We also obtain the correspondence between the weighted displacement and the
microrotations (\ref{lagrange multipliers}) and the kinematic variables by
applying (\ref{kin 1}) and (\ref{kin 2}) in integration of expressions (\ref%
{lagrange multipliers}):

\begin{eqnarray}
\Psi _{\alpha } &=&V_{\alpha }(x_{1},x_{2}),\text{ }W=w(x_{1},x_{2}),
\label{EQ correspondence} \\
\text{ }\Omega _{\alpha }^{0} &=&k_{1}\Theta _{\alpha }^{0}(x_{1},x_{2}),%
\text{ }\Omega _{3}=\frac{k_{2}}{h}\Theta _{3}(x_{1},x_{2}),  \notag \\
U_{\alpha } &=&U_{\alpha }(x_{1},x_{2}),\text{ }\Omega _{3}^{0}=\Theta
_{3}^{0}(x_{1},x_{2}),\   \notag
\end{eqnarray}%
where coefficients $k_{1\text{ }}$ and $k_{2\text{ }}$ depend on the
variation of microrotations. Under the conditions (\ref{kin 2}) we have that 
$k_{1}=\frac{4}{5}$ and $k_{2}=\frac{8}{5}.$

The density of the work over the boundary $\Gamma _{\sigma }\times \left[
-h/2,h/2\right] $

\begin{equation*}
\mathcal{W}_{2}=\frac{h}{2}\int_{-1}^{1}\left( \sigma _{o\alpha }u_{\alpha
}+m_{o\alpha }\varphi _{\alpha }\right) n_{\alpha }d\zeta _{3}
\end{equation*}%
can be presented in the form 
\begin{equation*}
\mathcal{W}_{2}=\mathcal{S}_{o}\mathcal{\cdot U=}\Pi _{o\alpha }\Psi
_{\alpha }+\Pi _{o3}W+M_{o\alpha }\Omega _{\alpha }^{0}+M_{o3}^{\ast }\Omega
_{3}+\Sigma _{o,\alpha }U_{\alpha }+\Upsilon _{o3}\Omega _{3}^{0},
\end{equation*}%
where 
\begin{eqnarray}
M_{\alpha \beta }n_{\beta } &=&\Pi _{o\alpha },\ R_{\alpha \beta }n_{\beta
}=M_{o\alpha },  \notag \\
Q_{\alpha }^{\ast }n_{\alpha } &=&\Pi _{o3},\ S_{\alpha }^{\ast }n_{\alpha
}=M_{o3}^{\ast }.
\end{eqnarray}

\begin{eqnarray}
N_{\alpha \beta }n_{\beta } &=&\Sigma _{\alpha }, \\
M_{\alpha }^{\ast }n_{\alpha } &=&\Upsilon _{o3}.
\end{eqnarray}%
Now $n_{\beta }$ is the outward unit normal vector to $\Gamma _{\sigma },$
and 
\begin{eqnarray}
\Pi _{o\alpha } &=&\left( \frac{h}{2}\right) ^{2}\int_{-1}^{1}\zeta
_{3}\sigma _{o\alpha }d\zeta _{3},\ M_{o\alpha }=\frac{h}{2}\int_{-1}^{1}\mu
_{o\alpha }d\zeta _{3},  \notag \\
\Pi _{o3} &=&\frac{h}{2}\int_{-1}^{1}\left( \sigma _{o3}-\sigma _{0}\right)
d\zeta _{3},\ M_{o3}^{\ast }=\frac{h}{2}\int_{-1}^{1}(\mu
_{o3}-tn_{3})d\zeta _{3},  \notag \\
\Sigma _{o,\alpha } &=&\frac{h}{2}\int_{-1}^{1}\sigma _{o\alpha }d\zeta
_{3},\ \Upsilon _{o3}=\left( \frac{h}{2}\right) ^{2}\int_{-1}^{1}\zeta
_{3}\left( \mu _{o3}-\zeta _{3}v\right) d\zeta _{3}.
\end{eqnarray}%
We are able to evaluate the work done at the top and bottom of the Cosserat
plate by\ using boundary conditions (\ref{Bound conditions 0}) and (\ref%
{Bound conditions 1a})

\begin{equation*}
\int_{T\cup B}\left( \sigma _{o3}u_{3}+m_{o3}\varphi _{o3}\right)
n_{3}da=\int\limits_{P_{0}}(pW+v\Omega _{3}^{0})da.
\end{equation*}

\subsection{The Cosserat plate internal work density}

Here we define the density of the work done by the stress and couple stress
over the Cosserat strain field:

\begin{equation}
\mathcal{W}_{3}=\frac{h}{2}\int_{-1}^{1}\left( \mathbf{\sigma \cdot \gamma
+\mu \cdot \chi }\right) d\zeta _{3}.  \label{C1}
\end{equation}

Substituting stress and couple stress assumptions (\ref{stress assumption 1}%
) - (\ref{stress assump 3(a)}) and integrating expression (\ref{C1}) \ we
obtain the following expression:

\begin{equation}
\mathcal{W}_{3}=\mathcal{S\cdot E=}M_{\alpha \beta }e_{\alpha \beta
}+Q_{\alpha }\omega _{\alpha }+Q_{3\alpha }^{\ast }\omega _{\alpha }^{\ast
}+R_{\alpha \beta }\tau _{\alpha \beta }+S_{\alpha }^{\ast }\tau _{3\alpha
}+N_{\alpha \beta }\upsilon _{\alpha \beta }+M_{\alpha }^{\ast }\tau
_{3,\alpha }^{0},  \label{Work density 2}
\end{equation}%
where$\ \mathcal{E}$ is the Cosserat plate strain set of the the weighted
averages of strain and torsion tensors%
\begin{equation*}
\ \mathcal{E=}\left[ e_{\alpha \beta },\omega _{\beta },\omega _{a}^{\ast
},\tau _{3\alpha },\tau _{\alpha \beta },\upsilon _{\alpha \beta },\tau
_{3,\alpha }^{0}\right] .
\end{equation*}%
Here the components of\ $\mathcal{E}$ are 
\begin{eqnarray}
e_{\alpha \beta } &=&\frac{3}{h}\int_{-1}^{1}\zeta _{3}\gamma _{\alpha \beta
}d\zeta _{3},  \label{s1} \\
\omega _{\alpha } &=&\frac{3}{4}\int_{-1}^{1}\gamma _{\alpha 3}\left(
1-\zeta ^{2}\right) d\zeta _{3},  \label{s2} \\
\omega _{\alpha }^{\ast } &=&\frac{3}{4}\int_{-1}^{1}\gamma _{3\alpha
}\left( 1-\zeta ^{2}\right) d\zeta _{3},  \label{s3} \\
\tau _{3\alpha } &=&\frac{3}{h}\int_{-1}^{1}\zeta _{3}\chi _{3\alpha }d\zeta
_{3},  \label{s4} \\
\tau _{\alpha \beta } &=&\frac{3}{4}\int_{-1}^{1}\chi _{\alpha \beta }\left(
1-\zeta ^{2}\right) d\zeta _{3},  \label{s5} \\
\upsilon _{\alpha \beta } &=&\frac{1}{2}\int_{-1}^{1}\gamma _{\alpha \beta
}d\zeta _{3},  \label{s6} \\
\tau _{3\alpha }^{0} &=&\frac{1}{2}\int_{-1}^{1}\chi _{3\alpha }d\zeta _{3}.
\label{s7}
\end{eqnarray}

The components of Cosserat plate strain (\ref{s1})-(\ref{s7}) can also be
represented in terms of the components of set $\mathcal{U}$ by the following
formulas:{\ }%
\begin{eqnarray}
e_{\alpha \beta } &=&\Psi _{\beta ,\alpha }+\varepsilon _{3\alpha \beta
}\Omega _{3},\text{ }  \notag \\
\omega _{\alpha } &=&\Psi _{\alpha }+\varepsilon _{3\alpha \beta }\Omega
_{\beta }^{0},  \notag \\
\omega _{\alpha }^{\ast } &=&W_{,\alpha }+\varepsilon _{3\alpha \beta
}\Omega _{\beta }^{0},  \notag \\
\tau _{3\alpha } &=&\Omega _{3,\alpha },  \label{strain -displ plate} \\
\tau _{\alpha \beta }^{0} &=&\Omega _{\beta ,\alpha }^{0},  \notag \\
\upsilon _{\alpha \beta } &=&U_{\beta ,\alpha }+\varepsilon _{3\alpha \beta
}\Omega _{3}^{0},  \notag \\
\tau _{3\alpha }^{0} &=&\Omega _{3,\alpha }^{0}.  \notag
\end{eqnarray}%
We call the relation\ (\ref{strain -displ plate}) the Cosserat plate
strain-displacement relation.

\section{Cosserat Plate HPR Principle}

It is natural now to reformulate HPR variational principle for the Cosserat
plate $P$. Let $\mathcal{A}$ denote the set of all admissible states that
satisfy the Cosserat plate strain-displacement relation (\ref{strain -displ
plate}) and let $\Theta $ be a HPR\ functional on $\mathcal{A}$ defined \ by

\begin{equation}
\Theta ({\large s)}=U_{K}^{S}-\int\limits_{P_{0}}(\mathcal{S\cdot E}%
-pW+v\Omega _{3}^{0})da+\int_{\Gamma _{\sigma }}\mathcal{S}_{n}\mathcal{%
\cdot }\left( \mathcal{U-U}_{o}\right) ds+\int_{\Gamma _{u}}\mathcal{S}_{o}%
\mathcal{\cdot U}ds,  \label{free_energy}
\end{equation}%
for every ${\large s}=\left[ \mathcal{U},\mathcal{E},\mathcal{S}\right] \in 
\mathcal{A}.$Then 
\begin{equation*}
\delta \Theta ({\large s})=0
\end{equation*}%
is equivalent to the following plate bending (A)\ and twisting (B)\ mixed
problems.

A. The bending equilibrium system of equations:

\begin{eqnarray}
M_{\alpha \beta ,\alpha }-Q_{\beta } &=&0,  \label{equilibrium_equations 1_A}
\\
Q_{a,\alpha }^{\ast }+p &=&0,  \label{equilibrium_equations 1_B} \\
R_{\alpha \beta ,\alpha }+\varepsilon _{3\beta \gamma }\left( Q_{\gamma
}^{\ast }-Q_{\gamma }\right) &=&0,  \label{equilibrium_equations 1_C} \\
S_{\alpha ,\alpha }^{\ast }+\epsilon _{3\beta \gamma }M_{\beta \gamma } &=&0,
\label{equilibrium_equations 1_D}
\end{eqnarray}%
with the resultant traction boundary conditions :

\begin{eqnarray}
M_{\alpha \beta }n_{\beta } &=&\Pi _{o\alpha },\ R_{\alpha \beta }n_{\beta
}=M_{o\alpha },  \label{bc_0} \\
Q_{\alpha }^{\ast }n_{\alpha } &=&\Pi _{o3},\ S_{\alpha }^{\ast }n_{\alpha
}=\Upsilon _{o3},  \label{bc_1}
\end{eqnarray}%
at the part $\Gamma _{\sigma }$ and the resultant displacement boundary
conditions

\begin{equation}
\Psi _{\alpha }=\Psi _{o\alpha },\text{ }W=W_{o},\text{ }\Omega _{\alpha
}^{0}=\Omega _{o\alpha }^{0},\text{ }\Omega _{3}=\Omega _{o3},  \label{bu_1}
\end{equation}%
at the part $\Gamma _{u}.$

The constitutive formulas: 
\begin{eqnarray}
e_{\alpha \alpha } &=&\frac{\partial {\large \Phi }}{\partial M_{\alpha
\alpha }}=\frac{12(\lambda +\mu )}{h^{3}\mu (3\lambda +2\mu )}M_{\alpha
\alpha }-  \label{Const1-2} \\
&&\left\vert \varepsilon _{\alpha \beta 3}\right\vert \frac{6\lambda }{%
h^{3}\mu (3\lambda +2\mu )}M_{\beta \beta }-\frac{3\lambda }{5h\mu (3\lambda
+2\mu )}p,
\end{eqnarray}

\begin{equation*}
e_{\alpha \beta }=\frac{\partial {\large \Phi }}{\partial M_{\alpha \beta }}=%
\frac{3(\mu _{c}+\mu )}{h^{3}\mu _{c}\mu }M_{\alpha \beta }+\frac{3(\mu
_{c}-\mu )}{h^{3}\mu _{c}\mu }M_{\beta \alpha },\text{ }\alpha \neq \beta
\end{equation*}

\begin{eqnarray}
\omega _{\alpha } &=&\frac{\partial {\large \Phi }}{\partial Q_{\alpha }}=%
\frac{3(\mu _{c}-\mu )}{10h\mu _{c}\mu }Q_{\alpha }^{\ast }+\frac{3(\mu
_{c}+\mu )}{10h\mu _{c}\mu }Q_{\alpha }, \\
\omega _{\alpha }^{\ast } &=&\frac{\partial {\large \Phi }}{\partial
Q_{\alpha }^{\ast }}=\frac{3(\mu _{c}-\mu )}{10h\mu _{c}\mu }Q_{\alpha }+%
\frac{3(\mu _{c}+\mu )}{10h\mu _{c}\mu }Q_{\alpha }^{\ast },  \notag
\end{eqnarray}%
\begin{eqnarray}
\tau _{\alpha \alpha }^{0} &=&\frac{\partial {\large \Phi }}{\partial
R_{\alpha \alpha }}=\frac{6(\beta +\gamma )}{5h\gamma (3\beta +2\gamma )}%
R_{\alpha \alpha }-  \label{Const1-3} \\
&&\left\vert \varepsilon _{\alpha \beta 3}\right\vert \frac{3\beta }{%
5h\gamma (3\beta +2\gamma )}R_{\beta \beta }-\frac{\beta }{2\gamma (3\beta
+2\gamma )}t,  \notag \\
&&\tau _{\alpha \beta }^{0}=\frac{\partial {\large \Phi }}{\partial R_{\beta
\alpha }}=\frac{3(\epsilon -\gamma )}{10h\gamma \epsilon }R_{\alpha \beta }+%
\frac{3(\gamma +\epsilon )}{10h\gamma \epsilon }R_{\beta \alpha },\text{ }%
\alpha \neq \beta  \notag
\end{eqnarray}

\begin{equation}
\tau _{3\alpha }=\frac{\partial {\large \Phi }}{\partial S_{\alpha }^{\ast }}%
=\frac{3(\gamma +\epsilon )}{h^{3}\gamma \epsilon }S_{\alpha }^{\ast }.
\label{Const1-4}
\end{equation}

B.\ The twisting equilibrium system of equations:

\begin{eqnarray}
N_{\alpha \beta ,\alpha } &=&0,  \label{equilibrium_equations 2 _A} \\
M_{\alpha ,\alpha }^{\ast }+\epsilon _{3\beta \gamma }N_{\beta \gamma }+v
&=&0,  \label{equilibrium_equations 2 _B}
\end{eqnarray}%
with the resultant traction boundary conditions at $\Gamma _{\sigma }$:

\begin{eqnarray}
N_{\alpha \beta }n_{\beta } &=&\Sigma _{\alpha },  \label{bc_2a} \\
M_{\alpha }^{\ast }n_{\alpha } &=&M_{o3}^{\ast },  \label{bc_2}
\end{eqnarray}%
and the resultant displacement boundary conditions at $\Gamma _{u}$:%
\begin{equation}
U_{\alpha }=U_{o\alpha },\text{ }\Omega _{3}^{0}=\Omega _{o3}^{0}.
\label{bu_2}
\end{equation}

The constitutive formulas:

\begin{eqnarray}
\omega _{\alpha \alpha } &=&\frac{\partial {\large \Phi }}{\partial
N_{\alpha \alpha }}=\frac{\lambda +\mu }{h\mu (3\lambda +2\mu )}N_{\alpha
\alpha }  \notag \\
&&-\frac{\lambda }{2h\mu (3\lambda +2\mu )}N_{(\alpha +1)(\alpha +1)}-\frac{%
\lambda }{2\mu (3\lambda +2\mu )}\sigma _{0},  \label{Const1-1} \\
\omega _{\alpha \beta } &=&\frac{\partial {\large \Phi }}{\partial N_{\alpha
\beta }}=\frac{\alpha +\mu }{4h\alpha \mu }N_{\alpha \beta }+\frac{\alpha
-\mu }{4h\alpha \mu }N_{\beta \alpha },\text{ }\alpha \neq \beta
\label{Const1-5} \\
\tau _{3\alpha }^{0} &=&\frac{\partial {\large \Phi }}{\partial M_{\alpha
}^{\ast }}=\frac{\gamma +\epsilon }{4h\gamma \epsilon }M_{\alpha }^{\ast }.
\label{Const1-6}
\end{eqnarray}%
We also represent the above constitutive relation in the compact form:%
\begin{equation*}
\mathcal{E=K}\left[ \mathcal{S}\right] =\mathcal{K\cdot S},
\end{equation*}%
where we call $\mathcal{K}$ the compliance Cosserat plate tensor.

Proof of the principle.\ The variation of $\Theta ({\large s)}$%
\begin{eqnarray*}
\delta \Theta ({\large s)} &=&\int\limits_{P_{0}}\left\{ \left( \mathcal{K}%
\left[ \mathcal{S}\right] -\mathcal{E}\right) \cdot \delta \mathcal{S}-%
\mathcal{S\delta E+}p\delta W+v\delta \Omega _{3}^{0}\right\} da \\
&&+\int_{\Gamma _{\sigma }}\left\{ \delta \mathcal{S}_{n}\mathcal{\cdot }%
\left( \mathcal{U-U}_{o}\right) +\mathcal{S}_{n}\mathcal{\cdot \delta U}%
\right\} ds+\int_{\Gamma _{u}}\mathcal{S}_{o}\mathcal{\cdot \delta U}ds.
\end{eqnarray*}

We apply Green's theorem and integration by parts for $\mathcal{S}$ and $%
\mathcal{\delta U}$ $\ \mathcal{\cite{Gurtin}}$ to the expression:

\begin{eqnarray*}
\int\limits_{P_{0}}\mathcal{S}\cdot \delta \mathcal{E}da
&=&\int\limits_{\partial P_{0}}\mathcal{S}_{o}\delta \mathcal{\cdot U}\text{ 
}ds\mathcal{-}\int\limits_{P_{0}}\{\left( M_{\alpha \beta ,\alpha }-Q_{\beta
}\right) \delta \Psi _{\beta }+Q_{\alpha ,\alpha }^{\ast }\delta W \\
&&+\left( R_{\alpha \beta ,\alpha }+\varepsilon _{3\beta \gamma }\left(
Q_{\gamma }^{\ast }-Q_{\gamma }\right) R_{\alpha \beta ,\alpha }\right)
\delta \Omega _{\beta }^{0} \\
&&+\left( S_{\alpha ,\alpha }^{\ast }+\epsilon _{3\beta \gamma }M_{\beta
\gamma }\right) \delta \Omega _{3}+N_{\alpha \beta ,\alpha }\delta U_{\beta }
\\
&&+\left( M_{\alpha ,\alpha }^{\ast }+\epsilon _{3\beta \gamma }N_{\beta
\gamma }\right) \delta \Omega _{3}^{0}\}da.
\end{eqnarray*}

Then based on the fact that $\mathcal{\delta U}$ and $\delta \mathcal{E}$
satisfy the Cosserat plate strain-displacement relation (\ref{strain -displ
plate}), we obtain

\begin{eqnarray*}
\delta \Theta ({\large s)} &=&\int\limits_{P_{0}}\left\{ \left( \mathcal{K}%
\left[ \mathcal{S}\right] -\mathcal{E}\right) \cdot \delta \mathcal{S}-%
\mathcal{S\delta E}\right\} da \\
&&+\int\limits_{P_{0}}\{\left( M_{\alpha \beta ,\alpha }-Q_{\beta }\right)
\delta \Psi _{\beta }+\left( Q_{\alpha ,\alpha }^{\ast }+p\right) \delta W \\
&&+\left( R_{\alpha \beta ,\alpha }+\varepsilon _{3\beta \gamma }\left(
Q_{\gamma }^{\ast }-Q_{\gamma }\right) R_{\alpha \beta ,\alpha }\right)
\delta \Omega _{\beta }^{0} \\
&&+\left( S_{\alpha ,\alpha }^{\ast }+\epsilon _{3\beta \gamma }M_{\beta
\gamma }\right) \delta \Omega _{3}+N_{\alpha \beta ,\alpha }\delta U_{\beta
}+\left( M_{\alpha ,\alpha }^{\ast }+\epsilon _{3\beta \gamma }N_{\beta
\gamma }+v\right) \delta \Omega _{3}^{0}\}da \\
&&+\int_{\Gamma _{\sigma }}\delta \mathcal{S}_{n}\mathcal{\cdot }\left( 
\mathcal{U-U}_{o}\right) ds+\int_{\Gamma _{u}}(\mathcal{S}_{o}-\mathcal{S}%
_{n})\mathcal{\cdot \delta U}ds.
\end{eqnarray*}

If ${\large s}$ is a solution of the mixed problem, then%
\begin{equation*}
\delta \Theta ({\large s)=}0.
\end{equation*}

On the other hand, some extensions of the fundamental lemma of calculus of
variations \cite{Gurtin} together with the fact that $\mathcal{U}$ and $%
\mathcal{E}$ satisfy the Cosserat plate strain-displacement relation (\ref%
{strain -displ plate}) imply that $\mathcal{S}$ is a solution of the A and B
mixed problems.

\textbf{Remark. } In the case of $\tau _{3\alpha }=\Omega _{3,\alpha }=0$ we
obtain that $S_{\alpha }^{\ast }=0.$ and $\mathbf{M}$ is symmetric, i.e. $%
M_{\alpha \beta }=M_{\beta \alpha }$ and the corresponding constitutive
relation is

\begin{equation*}
e_{\alpha \beta }=\frac{\partial {\large \Phi }}{\partial M_{\alpha \beta }}=%
\frac{6}{h^{3}\mu }M_{\alpha \beta }
\end{equation*}

The bending system for this case (\ref{equilibrium_equations 1_A})-(\ref%
{equilibrium_equations 1_D}) is reduced to the following:

\begin{eqnarray}
M_{\alpha \beta ,\alpha }-Q_{\beta } &=&0, \\
Q_{a,\alpha }^{\ast }+p &=&0, \\
R_{\alpha \beta ,\alpha }+\varepsilon _{3\beta \gamma }\left( Q_{\gamma
}^{\ast }-Q_{\gamma }\right) &=&0.
\end{eqnarray}

We also notice 
\begin{eqnarray*}
\omega -\omega _{\alpha }^{\ast } &=&\frac{3}{5h\mu _{c}}(Q_{\alpha
}-Q_{\alpha }^{\ast }) \\
\omega +\omega _{\alpha }^{\ast } &=&\frac{3}{5h\mu }(Q_{\alpha }+Q_{\alpha
}^{\ast })
\end{eqnarray*}%
and the case $\mu _{c}=0$ is consistent with the requirements 
\begin{equation}
Q_{\alpha }^{\ast }=Q_{\alpha }\text{ and }\Psi _{\alpha }=W_{,\alpha } 
\notag
\end{equation}%
Thus we obtain the equilibrium system in the decoupling form:%
\begin{eqnarray}
M_{\alpha \beta ,\alpha }-Q_{\beta } &=&0, \\
Q_{a,\alpha }+p &=&0, \\
R_{\alpha \beta ,\alpha } &=&0.
\end{eqnarray}

\section{Solution Uniqueness}

Here we prove that if there is a solution for the deformation of a Cosserat
elastic plate$,$ which satisfies the equilibrium equations (\ref%
{equilibrium_equations 1_A}) - (\ref{equilibrium_equations 1_D}), (\ref%
{equilibrium_equations 2 _A}) - (\ref{equilibrium_equations 2 _B}),
constitutive (\ref{Const1-2}) - (\ref{Const1-4}), (\ref{Const1-1}) - (\ref%
{Const1-6}) and kinematics formulas (\ref{strain -displ plate}) with
boundary conditions (\ref{bc_0}), (\ref{bc_1}), (\ref{bc_2}),(\ref{bc_2a})
at $\ \Gamma _{\sigma }$ and (\ref{bu_1}) and (\ref{bu_2}) at $\Gamma _{u}$
then this elastic solution must be unique. We also assume that all functions
and the plate middle plane region $\ P_{0}$ satisfy Green - Gauss theorem
requirements.

The proof will be based on contradiction. \ Let us assume that the solution
of the Cosserat plate is not unique in terms of the stresses and strains,
i.e. there would be two different solutions of (\ref{equilibrium_equations
1_A}) - (\ref{equilibrium_equations 1_D}) and (\ref{equilibrium_equations 2
_A}) - (\ref{equilibrium_equations 2 _B}), both of which satisfy the same
boundary conditions (\ref{Bound conditions 1}) and (\ref{Bound conditions 2}%
) at $\ \Gamma _{\sigma }$ and (\ref{bu_1}) and (\ref{bu_2}) at $\Gamma _{u}$%
. Due to linearity of the proposed \ model, the difference between these two
different solutions is also a solution of \ the same system of equations
with the following zero boundary conditions:%
\begin{eqnarray}
M_{\alpha \beta }n_{\beta } &=&0,\ R_{\alpha \beta }n_{\beta }=0,
\label{0 BC1} \\
Q_{\alpha }^{\ast }n_{\alpha } &=&0,\ S_{\alpha }^{\ast }n_{\alpha }=0, 
\notag
\end{eqnarray}%
\begin{equation}
N_{\alpha \beta }n_{\alpha }=0\ ,M_{\alpha }^{\ast }n_{\alpha }=0\ ,
\label{0 BC2a}
\end{equation}%
or 
\begin{equation}
\Psi _{\alpha }=0,W=0,\text{ }V\text{ }_{\beta }=0,\text{ }\Omega _{i}^{0}=0,%
\text{ }\Omega _{3}=0,\text{ }U_{\alpha }=0.  \label{0 BC3}
\end{equation}

It can be shown that for zero loads, the internal work ${\large U}$ can be
expressed by applying integration by parts as follows:

\begin{eqnarray}
{\large U} &=&\int\limits_{P_{0}}\mathcal{S\cdot E}da=\int\limits_{P_{0}}\{%
\left( M_{\alpha \beta }\Psi _{\alpha }+Q_{\beta }^{\ast }W+R_{\alpha \beta
}\Omega _{\alpha }^{0}+S_{\beta }^{\ast }\Omega _{3}+N_{\alpha \beta
}U_{\alpha }+M_{\beta }^{\ast }\Omega _{3}^{0}\right) _{,\beta }  \notag \\
&&-\left( M_{\alpha \beta ,\alpha }-Q_{\beta }\right) \Psi _{\beta
}-Q_{\alpha ,\alpha }^{\ast }W  \notag \\
&&-\left( R_{\alpha \beta ,\alpha }+\varepsilon _{3\beta \gamma }\left(
Q_{\gamma }^{\ast }-Q_{\gamma }\right) R_{\alpha \beta ,\alpha }\right)
\Omega _{\beta }^{0}  \label{uniqueness 1A} \\
&&-\left( S_{\alpha ,\alpha }^{\ast }+\epsilon _{3\beta \gamma }M_{\beta
\gamma }\right) \Omega _{3}-N_{\alpha \beta ,\alpha }U_{\beta }  \notag \\
&&-\left( M_{\alpha ,\alpha }^{\ast }+\epsilon _{3\beta \gamma }N_{\beta
\gamma }\right) \Omega _{3}^{0}\}da.  \notag
\end{eqnarray}%
Taking into account Green's theorem, the equilibrium equations (\ref%
{equilibrium_equations 1_A}) - (\ref{equilibrium_equations 1_D}) and (\ref%
{equilibrium_equations 2 _A}) - (\ref{equilibrium_equations 2 _B}), \
expression (\ref{uniqueness 1A}) is reduced to the following line integral: 
\begin{equation}
{\large U}=\oint_{\Gamma }\mathcal{S}_{n}\mathcal{\cdot U}ds=\oint_{\Gamma
}\left( \check{M}_{\alpha }\Psi _{\alpha }+\check{Q}^{\ast }W+\check{R}%
_{\alpha }\Omega _{\alpha }^{0}+\check{S}^{\ast }\Omega _{3}+\check{N}%
_{\alpha }U_{\alpha }+\check{M}^{\ast }\Omega _{3}^{0}\right) ds=0,
\label{uniqueness 1B}
\end{equation}%
which vanishes because of the zero boundary conditions (\ref{0 BC1})-(\ref{0
BC3}).

Using the constitutive equation (\ref{Const1-2}) - (\ref{Const1-4}), (\ref%
{Const1-1}) - (\ref{Const1-6}) in a reversible form, the positive definite
quadratic form strain energy density (\ref{Energy density}) can be
represented in terms of the Cosserat plate strain set $\ \mathcal{E}$, which
components in this case should be zeros. Then from the Cosserat plate
strain-displacement relation (\ref{strain -displ plate}) we obtain the
system:

\begin{eqnarray*}
U_{1,1} &=&0,\text{ }U_{2,2}=0,\text{ }U_{2,1}-\Omega _{3}^{0}=0,\text{ }%
U_{1,2}+\Omega _{3}^{0}=0, \\
\text{ }W_{,1}+\Omega _{2}^{0} &=&0,\text{ }W_{,2}-\Omega _{1}^{0}=0,\text{ }%
\Psi _{1,1}=0,\text{ }\Psi _{2,2}=0,\text{ }\Omega _{3,1}=0,\text{ }\Omega
_{3,2}=0, \\
\Psi _{2,1}-\Omega _{3} &=&0,\text{ }\Psi _{1,2}+\Omega _{3}=0,\text{ }\Psi
_{1}-\Omega _{2}^{0}=0,\text{ }\Psi _{2}+\Omega _{1}^{0}=0, \\
\Omega _{1,1}^{0} &=&0,\text{ }\Omega _{2,2}^{0}=0,\text{ }\Omega
_{1,2}^{0}=0,\text{ }\Omega _{2,1}^{0}=0,\text{ }\Omega _{3,1}^{0}=0,\text{ }%
\Omega _{3,2}^{0}=0,
\end{eqnarray*}%
which has the following solution: 
\begin{equation*}
U_{\alpha }=\varepsilon _{3\beta \alpha }\Omega _{3}^{0}x_{\beta }+U_{\alpha
}^{0},\text{ }W=\varepsilon _{3\alpha \beta }\Omega _{\alpha }^{0}x_{\beta
}+W^{0},\text{ }\Psi _{\alpha }=\varepsilon _{3\alpha \beta }\Omega _{\beta
}^{0},\text{ }\Omega _{3}=0,
\end{equation*}%
\ \ where constant parameters $U_{\alpha }^{0},$ $W^{0}$ are the rigid
translations of the middle plane, $\Omega _{i}^{0}$ the component of the
rigid rotation of the plane around $x_{i}$ axis, and $\Omega _{3}$ the slope
of the rigid rotation around $x_{3}.$ Thus the difference between any two
deformations and microrotations of the plate, having the same boundary
conditions, represents changes of the plate as a rigid body.

\section{Conclusion}

We generalized Hellinger-Prange -Reissner (HPR)\ principle in order to
derive the new equilibrium equations in the middle plane and constitutive
relationships for the plate. The polynomial approximations of the variation
of couple stress and micropolar rotations in the thickness direction in
order higher than one allowed us, based on the generalized HPR\ principle,
to project Cosserat 3D\ Elasticity equilibrium equations into the new form
of equilibrium equations and the constitutive relations in the middle plane
of the plate.

\section{Acknowledgement}

I express my gratitude\ to Krzysztof R\'{o}zga and Darrell Hajek\ for
providing assistance in the presentation of the paper.

\end{document}